\documentclass[aps,reprint,amsmath,amssymb,showpacs,superscriptaddress]{revtex4-1}
\usepackage{graphicx}
\usepackage{SIunits}
\usepackage{bm}     
\usepackage{hyperref} 
\usepackage{dsfont}    
\usepackage{color}
\DeclareMathOperator{\Tr}{Tr}

\begin{document}

\title{Experimental comparison of various quantum key distribution protocols \\under reference frame rotation and fluctuation}

\author{Juhwan Yoon}
\affiliation{Center for Quantum Information, Korea Institute of Science and Technology (KIST), Seoul, 02792, Republic of Korea} 
\affiliation{Department of Electrical and Computer Engineering, Ajou University, Suwon, 16499, Republic of Korea}

\author{Tanumoy Pramanik}
\affiliation{Center for Quantum Information, Korea Institute of Science and Technology (KIST), Seoul, 02792, Republic of Korea}

\author{Byung Kwon Park}
\affiliation{Center for Quantum Information, Korea Institute of Science and Technology (KIST), Seoul, 02792, Republic of Korea}
\affiliation{Division of Nano \& Information Technology, KIST School, Korea University of Science and Technology, Seoul 02792, Republic of Korea}

\author{Sang-Wook Han}
\affiliation{Center for Quantum Information, Korea Institute of Science and Technology (KIST), Seoul, 02792, Republic of Korea} 

\author{Sangin Kim}
\affiliation{Department of Electrical and Computer Engineering, Ajou University, Suwon, 16499, Republic of Korea}

\author{Yong-Su Kim}\email{yong-su.kim@kist.re.kr}
\affiliation{Center for Quantum Information, Korea Institute of Science and Technology (KIST), Seoul, 02792, Republic of Korea}
\affiliation{Division of Nano \& Information Technology, KIST School, Korea University of Science and Technology, Seoul 02792, Republic of Korea}

\author{Sung Moon}
\affiliation{Center for Quantum Information, Korea Institute of Science and Technology (KIST), Seoul, 02792, Republic of Korea}
\affiliation{Division of Nano \& Information Technology, KIST School, Korea University of Science and Technology, Seoul 02792, Republic of Korea}

\date{\today} 

\begin{abstract}
\noindent	
In free-space quantum key distribution (QKD) between moving parties, e.g., free-space QKD via satellite, the reference frame rotation and fluctuation degrades the performance of QKD. Reference-frame-independent QKD (RFI-QKD) provides a simple but efficient way to overcome this problem. While there has been a number of theoretical and experimental studies on RFI-QKD, the experimental verification of the robustness of RFI-QKD over other QKD protocols under the reference frame rotation and fluctuation is still missing. Here, we have constructed a free-space QKD system which can implement BB84, six-state, and RFI-QKD protocols, and compared their performances under the reference frame rotation and fluctuation. With the theoretical analysis and experimental data, we have successfully verified the robustness of RFI-QKD protocol over other QKD protocols in the presence of reference frame rotation and fluctuation.
\end{abstract}

\maketitle

\section{Introduction}
Quantum key distribution (QKD), one of the most mature quantum information technologies, provides a way to distribute the secrete keys between distant parties of Alice and Bob with the security being guaranteed by the laws of quantum physics~\cite{Bennett84,Ekert91}. There has been enormous effort to improve the security and practicality of the QKD both in theory and experiment~\cite{hwang03,lydersen10,Lo12,Guan15,Choi16,Lee17}. Recently, satellite based free-space QKD has been attracted a lot of attention since it overcomes the communication distance limit of optical fiber based QKD~\cite{takenaka17,liao17,ren17,yin17,bedington17}. Indeed, it has been recently reported that the intercontinental QKD communication is possible via satellite quantum communication~\cite{liao18}. Note that, in free-space QKD, polarization is the most commonly chosen degree of freedom for qubit encoding due to the robustness  and simplicity for implementation.

In order to implement QKD, it is usually required to share a common reference frame between Alice and Bob. In free-space QKD, the polarization axes should be aligned and maintained during the communication. However, in satellite QKD, it can be difficult and costly to maintain the polarization axes alignment due to the revolution and rotation of the satellite with respect to the ground station~\cite{Yin13}. Reference-frame-independent (RFI) QKD provides an efficient way to solve this problem~\cite{Laing10,Wabnig13,Zhang14,Liang14}. Note that the hardware requirement for RFI-QKD is comparable to that of BB84 or six-state protocols, the post processing using classical communication is different from others. In RFI-QKD protocol, the secret keys are shared via the basis which is not affected by the reference frame alignment while the security is checked by other two non-commuting bases. Note that the RFI-QKD can be applied to measurement-device-independent QKD~\cite{Wang15,Zhang17}.

In the original proposal and following experimental implementation, the security of RFI-QKD has been discussed only when the reference frames of Alice and Bob are rotated with a fixed angle, so there is no relative change between the reference frames during the communication~\cite{Laing10,Wabnig13,Zhang14,Liang14}. In practice, however, the reference frame of satellite with respect to the ground station continuously changes. There has been only few theoretical studies to analyze the security of RFI-QKD under the continuously changing reference frames~\cite{Sheridan10,Pramanik17}, however, there has been no experimental verification. 

In this paper, we have both theoretically and experimentally investigated various QKD protocols including BB84, six-state, and RFI-QKD protocols under the reference frame rotation and fluctuation. By comparing the secret key rates of the QKD protocols, we have verified that RFI-QKD indeed outperforms other QKD protocols under the continuously changing reference frames.

\begin{figure}[b]
\includegraphics[width=3.3in]{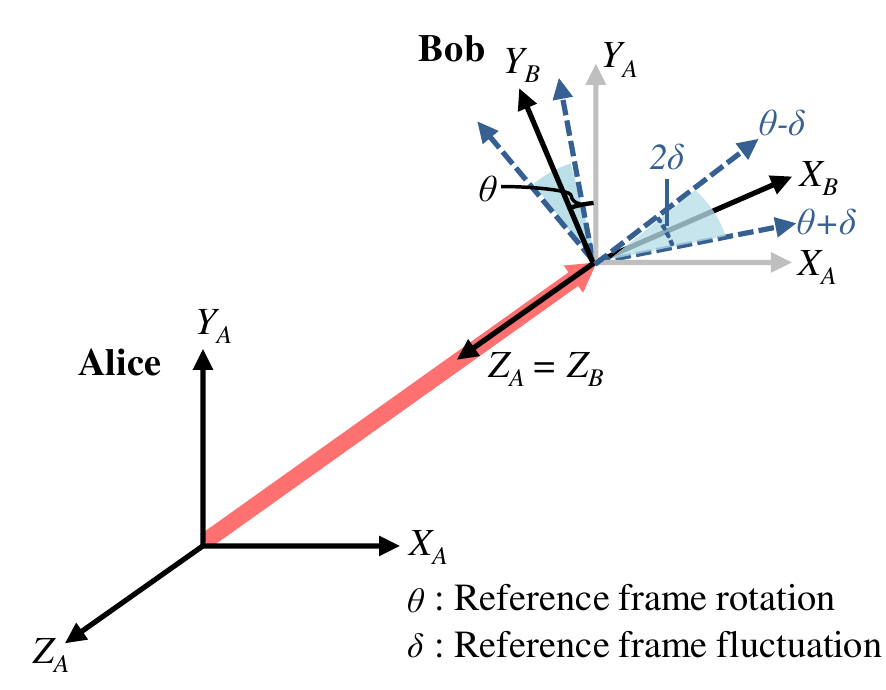}
\caption{The reference frames shared by Alice and Bob. $Z$ axes of Alice and Bob are always perfectly aligned as long as they face each other.}
\label{scheme}
\end{figure}

\section{Theory}

Let us suppose that Alice prepares a single-photon state in the polarization state $\rho_A$ either along $X$ or $Y$ or $Z$ bases and sends it to Bob. Here, $X$, $Y$, and $Z$ bases form a mutually unbiased bases in a single qubit or a 2 dimensional Hilbert space. It is often to define $X$, $Y$, and $Z$ bases as $H/V$, $D/A$, and $R/L$ bases, where H, V, D, A, R, and L denote horizontal, vertical, diagonal (45$^\circ$), anti-diagonal (-45$^\circ$), right circular, and left circular polarization states, respectively. Assuming the isotropic transmission channel noise $p$, Bob receives the state $\rho_B$ of
\begin{eqnarray}
\rho_B= (1-p) \, \rho_A +p\,\frac{I}{2},
\label{state}
\end{eqnarray}
where $I$ is the identity matrix.

\begin{figure*}[t]
\includegraphics[width=7in]{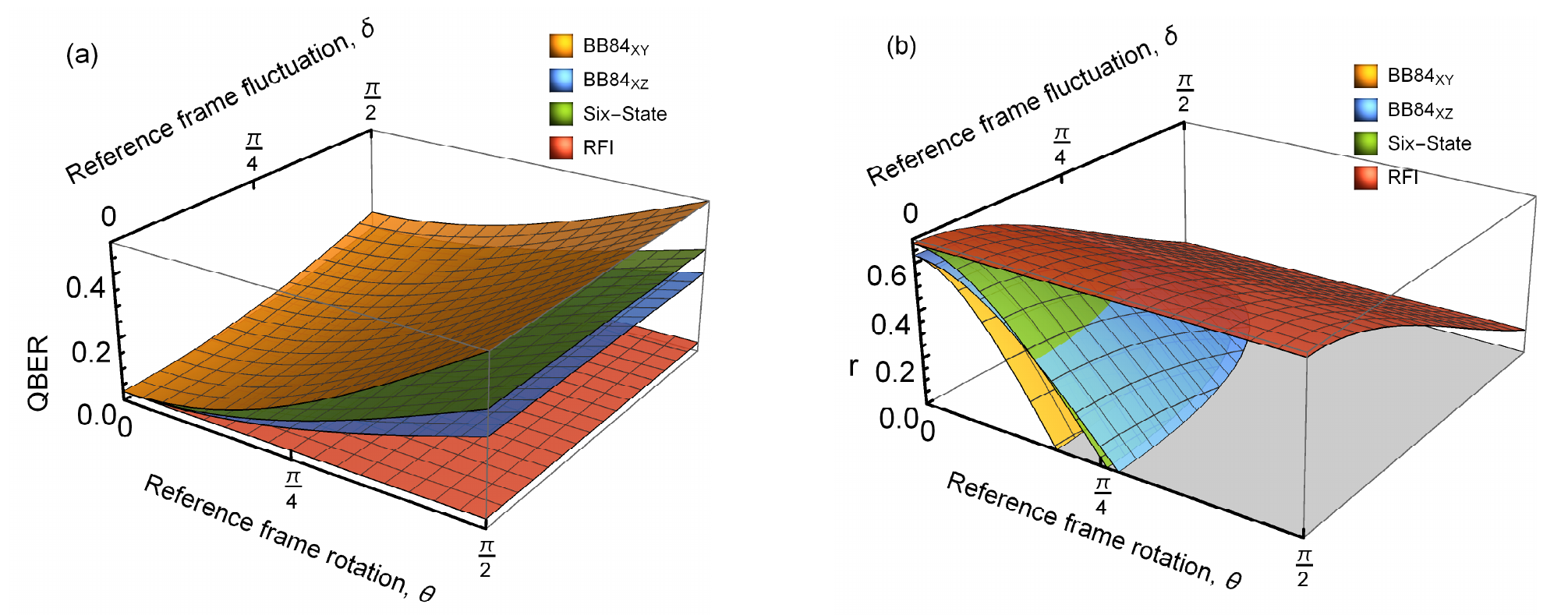}
\caption{Theoretical (a) QBER and (b) secret key rate $r$ of various QKD protocols with respect to the reference frame rotation $\theta$ and fluctuation $\delta$. The channel noise is assumed to be $p=0.06$, so $Q_Z=3~\%$.}\label{theory}
\end{figure*}
 
Figure~\ref{scheme} presents the polarization axes of Alice is fixed while that of Bob is rotated with the angle of $\theta$ with respect to the Alice's polarization axis. Bob's polarization axis also fluctuates with the amount of $\pm\delta$. Therefore, during the communication, Bob's $X$ and $Y$ axes with respect to those of Alice changes from $\theta-\delta$ to $\theta+\delta$. At a certain time, the polarization axes of Alice and Bob are correlated as
\begin{eqnarray}
X_B &=& X_A \, \cos{\phi}+  Y_A \, \sin{\phi},  \nonumber\\
Y_B &=& Y_A \, \cos{\phi}-  X_A \, \sin{\phi},   \nonumber\\
Z_B &=& Z_A,
\label{Rot_Fram}
\end{eqnarray}
where the subscripts $A$ and $B$ denote Alice and Bob, and $\phi$ takes values from the range of $\theta-\delta\leq\phi\leq\theta+\delta$. Note that the $Z$ axes of Alice and Bob are always perfectly aligned as long as they face each other.

The quantum bit error rate (QBER) in each basis is given as
\begin{eqnarray}
Q_W=\frac{1-\overline{\langle W_A\, W_B\rangle}}{2},
\label{QBER}
\end{eqnarray}
where $W\in\{X,\,Y,\,Z\}$. Here, the average $\overline{\langle W_A\, W_B\rangle}$ is taken over the fluctuation of $\pm\delta$ as
\begin{eqnarray}
\overline{\langle W_A\,W_B\rangle} &=& \frac{1}{2\delta} \int_{\theta-\delta}^{\theta+\delta} \langle W_A\,W_B\rangle d\phi \nonumber\\
&=& \langle W_A\,W_B\rangle \, \cos\theta \, \frac{\sin\delta}{\delta},
\label{Cor_av}
\end{eqnarray}
where the expectation value $\langle W_A\,W_B\rangle$ can be calculated as
\begin{eqnarray}
\langle W_A\,W_B\rangle = \Tr[(W_A \otimes W_B) \cdot (\rho_A \otimes \rho_B)].
\end{eqnarray}
In the present scenario, the QBER in $X$, $Y$ and $Z$ bases are given by
\begin{eqnarray}
Q_{X}[\theta,\, \delta]  &=& Q_{Y}[{\theta}, {\delta}]= \frac{1}{2}-\frac{(1- Q_{Z} ) \mathrm{cos {\theta} \ sin {\delta}}}{2{\delta}},\nonumber\\
Q_{Z}[\theta,\, \delta] &=&Q_Z = \frac{p}{2}. 
\label{QZX}
\end{eqnarray}
It is remarkable that $Q_Z$ is independent of the reference frame rotation and fluctuation since $Z$ axis is well aligned all the time.

In BB84 protocol, the secrete keys are established via two non-commuting bases. If Alice and Bob use $X$ and $Y$ bases, the average QBER becomes  
\begin{eqnarray}
Q_{\text{BB84}}^{XY}[\theta,\, \delta]  &=& \frac{Q_X[\theta,\, \delta] \,+\,Q_Y[\theta,\, \delta] }{2}.
\end{eqnarray}
Since $Z$ basis is invariant under the reference frame rotation and fluctuation, Alice and Bob can expect lower QBER by using $X$ and $Z$ bases for BB84 QKD protocol as
\begin{eqnarray}
Q_{\text{BB84}}^{XZ}[\theta,\, \delta]  &=& \frac{Q_X[\theta,\, \delta] \,+\,Q_Z[\theta,\, \delta] }{2}.
\end{eqnarray}
Six-state protocol utilizes all three mutually unbiased bases for secrete key generation, so the average QBER is
\begin{eqnarray}
Q_{\text{six}}^{XYZ}[\theta,\, \delta]  &=& \frac{Q_X[\theta,\, \delta] \,+\,Q_Y[\theta,\, \delta]  + \, Q_Z[\theta,\, \delta]}{3}.
\end{eqnarray}
The secret key rates of BB84 and six-state protocol can be calculated with QBER as~\cite{Bruss98,Lo01}
\begin{eqnarray}
r^{XY(XZ)}_{\text{BB84}} &=& 1-2H[Q^{XZ(XY)}_{\text{BB84}}], \nonumber \\
r^{XYZ}_{\text{six}} &=& 1-H[Q_{\text{six}}^{XYZ}]-Q_{\text{six}}^{XYZ}\nonumber\\
&&-(1-Q_{\text{six}}^{XYZ})H\left [\frac{1-\frac{3}{2}Q_{\text{six}}^{XYZ}}{1-Q_{\text{six}}^{XYZ}} \right],
\label{r_BB84_6}
\end{eqnarray}
where $H[x] = -x\log_{2}x-(1-x)\log_{2}(1-x)$ is the Shannon entropy.

In RFI-QKD protocol, the secrete key is obtained from the polarization states along $Z$-axis which is unaffected by the reference frame rotation and fluctuation while the correlation in other two bases is used to bound the knowledge of an eavesdropper. To this end, a parameter $C$ is defined as~\cite{Laing10},
\begin{eqnarray}
C= \langle X_A X_B \rangle^2 +\langle X_A Y_B \rangle^2 + \langle Y_A X_B \rangle^2 + \langle Y_A Y_B \rangle^2 .
\label{CTh}
\end{eqnarray}
Under the reference frame rotation and fluctuation, the average of $C$ becomes~\cite{Pramanik17}
\begin{eqnarray}
C[\theta,\, \delta] = 2\left(1-2Q_{Z}\right)^2 \left(\frac{\sin {\delta}}{\delta}\right)^2,
\label{CValue}
\end{eqnarray}
where the average is taken over fluctuation of reference frame. Note that $C$ parameter is independent of the reference frame rotation $\theta$, however, the reference frame fluctuation $\delta$ affects to $C$ parameter. 

Using the QBER and $C$ parameter, one can estimate the Eve's knowledge as~\cite{Laing10} 
\begin{eqnarray}
I_E[Q_{Z},C] =  (1 - Q_{Z}) H \left[\frac{1+u}{2}\right] + Q_{Z} H\left[\frac{1+v}{2}\right],
\label{IE}
\end{eqnarray}
where 
\begin{eqnarray}
u  &=& \min\left [ \frac{1}{1-Q_{Z}} \sqrt{\frac{C[\theta,\, \delta]}{2}},1\right] \nonumber \\
v  &=& \frac{1}{Q_{Z}} \sqrt{\frac{C[\theta,\, \delta]}{2}-(1-Q_{Z})^2 \,u^2}. 
\end{eqnarray}
The secret key rate in RFI-QKD protocol can be calculated using Eqs.~(\ref{QZX}), (\ref{CValue}) and (\ref{IE}), and it becomes~\cite{Laing10},
\begin{eqnarray}
r_{{\rm RFI}} = 1-H[Q_{Z}]-I_E[Q_{Z},C]
\label{r_RFI}
\end{eqnarray}

Figure~\ref{theory} (a) and (b) present the QBER and the secret key rates for various QKD protocols with respect to the reference frame rotation and fluctuation, respectively. The orange, blue, green and red surfaces correspond to BB84 with $\{X,Y\}$, and $\{X,Z\}$ bases, six-state, and RFI-QKD protocols, respectively. Here, we assume the channel noise $p=0.06$, and thus $Q_Z=3~\%$. While the QBER for BB84 with $\{X,Y\}$ bases, $Q_{\text{BB84}}^{XY}$, is the most sensitive with respect to the reference frame rotation and fluctuation, that of RFI-QKD, $Q_Z$, is invariant. Figure~\ref{theory} (b) also shows that the secret key rate of RFI-QKD is unaffected by the reference frame rotation $\theta$ and decreases with strength of fluctuation $\delta$, while the secrete key rates for other QKD protocols are affected by both reference frame rotation and fluctuation. It also shows that RFI-QKD provides more secret keys than other QKD protocols under the reference frame rotation and fluctuation. 

\begin{figure*}[t]
\includegraphics[width=6.7in]{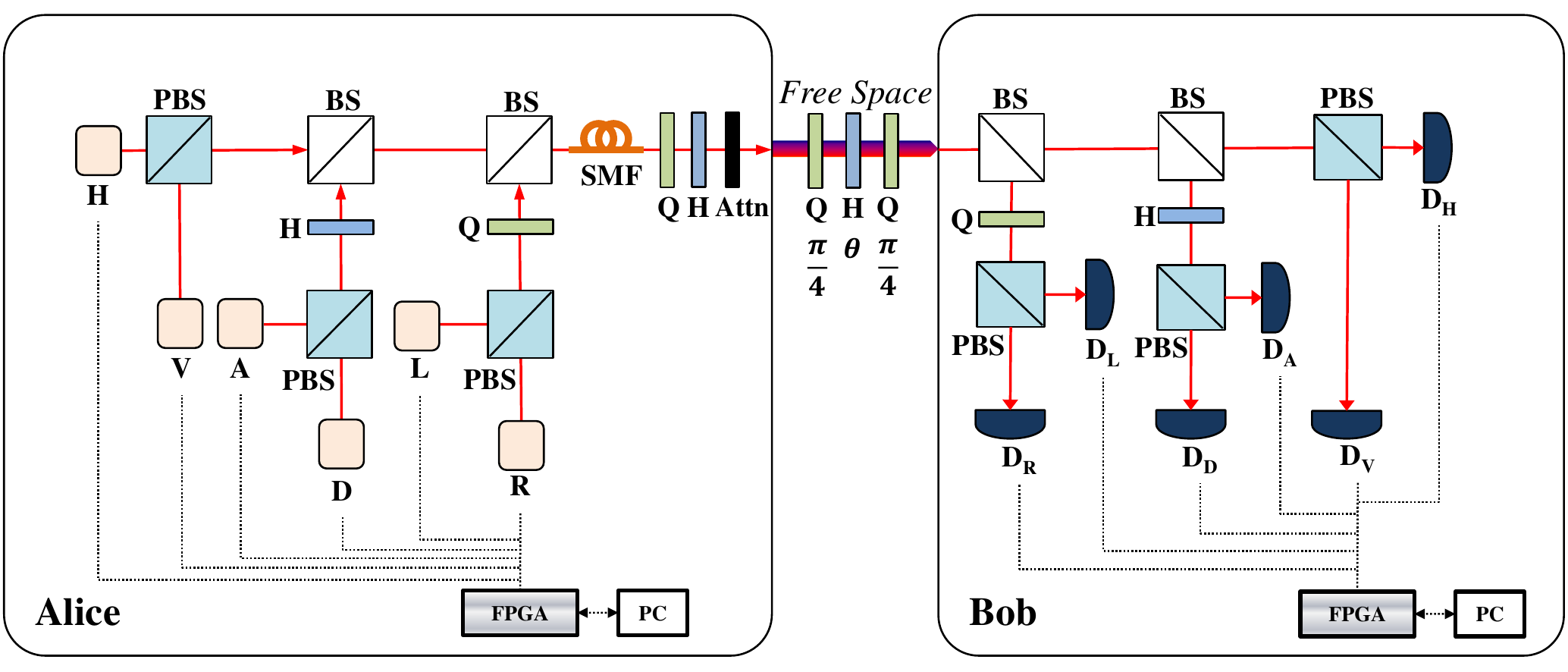}
\caption{Experimental setup for RFI-QKD. BS: beamsplitter, H: half waveplate, Q: quarter waveplate, PBS: polarizing beamsplitter, SMF: single-mode fiber, Attn: attenuator, D: singe-photon detector. The whole system is controlled FPGA boards and computers via a homemade graphical user interface (GUI). In free-space channel, quarter-half-quarter waveplates are placed to preciesely demonstrate the reference frame rotation and fluctuation.}\label{setup}
\end{figure*}

\section{Experiment}

\subsection{Experimental setup}

The experimental setup to investigate various QKD protocols under the reference frame rotation and fluctuation is depicted in Fig.~\ref{setup}. Alice prepares coherent pulses with the polarization state of either $|H\rangle, |V\rangle, |D\rangle, |A\rangle, |R\rangle$ or $|L\rangle$ using six identical lasers and linear optical elements. The wavelength and pulse width of the coherent pulses are 780~nm and 3~ns, respectively. The repetition rate of the system is 100~MHz. The coherent pulses from different lasers pass through a single-mode fiber (SMF) for spatial mode filtering. Then, waveplates (HWP and QWP) are used to compensate the polarization mode dispersion during the single-mode fiber transmission. An optical attenuator is placed at the end of the transmitter to prepare weak coherent pulses with appropriate mean photon number per pulse of $\mu\approx0.5$. Bob uses six single-photon detectors ${\rm D}_P$ where the subscript $P$ is the polarization state to detect weak coherent pulses. 

Alice and Bob use field programmable gate array (FPGA) boards and computers (PC) to perform the necessary system control including random number generation and data processing. The transceiver in the FPGA board controls the laser operation and the key sifting process. All the information about control and detection of transmitter and receiver are stored in the internal memory of the FPGA and transmitted over ethernet communications between FPGA and PC. The whole operation is controlled by a homemade graphical user interface (GUI) at a computer which is programmed in C\#. The classical communication between Alice and Bob for key sifting, error correction, and privacy amplification is performed via long range wireless communication channel so as to implement the whole classical and quantum communication via free-space.
 

In order to precisely implement the polarization axes rotation and fluctuation, we use a set of waveplates in the free-space channel. By rotating a HWP which is placed between two QWP at $\pi/4$, one can implement the polarization axes rotation. The transformation of QWP($\pi/4$)-HWP($\theta_H$)-QWP($\pi/4$) is presented by the unitary matrix of
\begin{eqnarray}
U = Q\left(\frac{\pi}{4}\right) H\left(\theta_H\right) Q\left(\frac{\pi}{4}\right) = \begin{bmatrix} e^{-i2 \theta _H} & 0 \\ 0 & -e^{i2 \theta _H} \end{bmatrix}.
\label{U_rot}
\end{eqnarray}
Assuming the polarization states $\{D,\,A\}$, $\{R,\,L\}$, $\{H,\,V\}$ as $X, Y$, and $Z$ bases, respectively, the effect of $U$ is given as 
\begin{eqnarray}
{W}_{B} &=&U{W}_{A}U^{\dagger}  
\label{UXY}
\end{eqnarray}
where $W\in\{X,Y,Z\}$. By putting $\phi=4\,\theta_H$, one can realize the reference frame rotation of Eq.~(\ref{Rot_Fram}).

\begin{figure*}[t]
\includegraphics[width=7in]{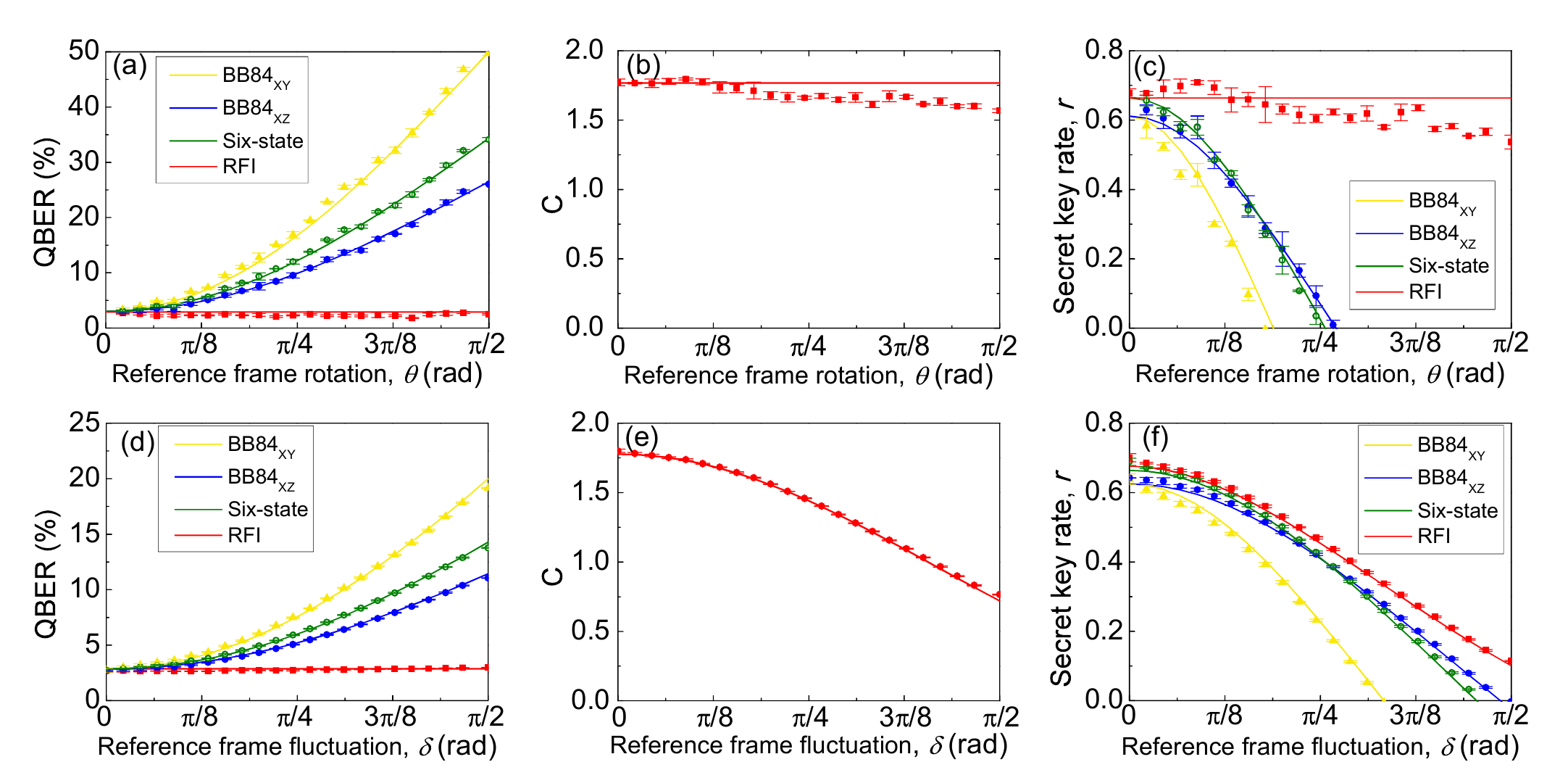}
\caption{(a), (b) and (c) are QBER, $C$ parameter, and secret key rates under the fixed reference frame rotation $\theta$ with $\delta=0$. (d), (e) and (f) are QBER, $C$ parameter, and secret key rates under the reference frame fluctuation $\delta$ without reference frame rotation $\theta=0$. In both scenarios, RFI-QKD shows the highest secret key rate. Error bars are experimentally obtained standard deviations. 
}\label{data}
\end{figure*}

\subsection{Experimental result}

Figure~\ref{data}(a), (b), and (c) show QBER, $C$ parameter, and the secret key rates with respect to the fixed reference frame rotation, i.e., $0\le\theta$ while $\delta=0$, respectively. The yellow, blue, green and red marks (lines) correspond experimental (theoretical) values for BB84 with $\{X,\,Y\}$ bases, BB84 with $\{X,\,Z\}$ bases, six-state and RFI-QKD protocols, respectively. Figure~\ref{data}(a) shows that QBER for BB84 protocol with $\{X,\,Y\}$ bases is larger than those of other QKD protocols. In particular, QBER for RFI-QKD is unaffected under the reference frame rotation. The non-vanishing QBER of $Q_z=3~\%$ in RFI-QKD occurs due to the channel noise and device imperfection. As shown in Fig.~\ref{data}(b), the $C$ parameter is also independent of the reference frame rotation. Figure~\ref{data}(c) shows that the secret key rates of all other QKD protocols decreases as the reference frame rotation $\theta$ increases. In particular, BB84 with $\{X,\,Y\}$ bases, BB84 with $\{X,\,Z\}$ bases, and six-state protocols fail to obtain secrete keys when the reference frame rotation is close to the theoretical values of $\theta\sim0.19\pi$, $0.27\pi$, and $0.26\pi$, respectively. On the other hand, the secret key rate in RFI-QKD protocol is independent of the reference frame rotation, and thus, outperforms other QKD protocols.

The effects of the reference frame fluctuation $0\le\delta$ on QBER, $C$ parameter, and secret key rates are presented in Fig.~\ref{data}(d), (e), and (f), respectively. In order to rule out the effect of the reference frame rotation, we set $\theta=0$. To simulate the effect of the reference frame fluctuation, the experimental data are taken for fixed reference frame rotation $\theta$, and statistically mixed. As shown in Fig.~\ref{data}(d) and (e), QBER of RFI-QKD is not affected by the reference frame fluctuation, but $C$ parameter decreases as the fluctuation increases. As a result, as shown Fig.~\ref{data}(f), the secret key rate of RFI-QKD protocol decreases with the fluctuation, similar to the other QKD protocols. However, apparently, RFI-QKD provides the best robustness against the reference frame fluctuation among various QKD protocols. Remarkably, RFI-QKD protocol provides non-zero secret key rate even when the reference frame fluctuates from $-\pi/2$ to $\pi/2$, i.e., the fluctuation induces the bases changes between $X$ and $Y$ where all other QKD protocols fail to provide secret keys. In the reference frame fluctuation scenario, the secret key rate under the BB84 protocol with $\{X,\,Y\}$ bases, BB84 protocol with $\{X,\,Z\}$ bases, six-state protocols vanishes at the strengths of fluctuation of close to their theoretical values of $\delta\sim0.33\pi$, $0.48\pi$, and $0.46\pi$, respectively.

\section{Conclusion}

In summary, we have experimentally demonstrated various QKD protocols of BB84, six-state, and RFI-QKD protocols under the reference frame rotation and fluctuation. In both cases, QBER of RFI-QKD protocol is invariant while that of other QKD protocols increases as the amount of reference frame rotation and fluctuation increases. With QBER and $C$ parameter of RFI-QKD, we have verified that RFI-QKD protocol provides the highest secrete key rate among the employed QKD protocols under the reference frame rotation and fluctuation. These theoretical and experimental results indicate that, in practical earth-satellite communication scenario where the polarization axis at the satellite continuously changes with respect to that on earth, RFI-QKD protocol can be a simple but powerful solution for a better QKD communication.

\section{Acknowledgments}
This work was supported by the ICT R$\&$D program of MSIP/IITP (B0101-16-1355), and the KIST research program (2E27801).

\end{document}